\documentclass[pra,twocolumn,showpacs,superscriptaddress,floatfix,nofootinbib]{revtex4-1}
\usepackage{epsfig}
\usepackage{latexsym}
\usepackage{amsmath}
\usepackage{graphics}

\usepackage{amsfonts}
\usepackage{float}

\usepackage{color}
\usepackage{pstricks}
\newcommand{\br}{\mathbf{r}}

\begin{document}

\title{Exact Casimir-Polder potentials:\\ interaction of an atom with a conductor-patched dielectric surface}
\author{Claudia Eberlein}
\author{Robert Zietal}
\affiliation{Department of Physics \& Astronomy,
    University of Sussex,
     Falmer, Brighton BN1 9QH, England}
\date{\today}
\begin{abstract}
We study the interaction between a neutral atom or molecule and a conductor-patched dielectric surface. We model this system by a perfectly reflecting disc lying atop of a non-dispersive dielectric half-space, both interacting with the neutral atom or molecule. We assume the interaction to be non-retarded and at zero temperature. We find an exact solution to this problem. In addition we generate a number of other useful results. For the case of no substrate we obtain the exact formula for the van der Waals interaction energy of an atom near a perfectly conducting disc. We show that the Casimir-Polder force acting on an atom that is polarized in the direction normal to the surface of the disc displays intricate behaviour. This part of our results is directly relevant to recent matter-wave experiments in which cold molecules are scattered by a radially symmetric object in order to study diffraction patterns and the so-called Poisson spot. Furthermore, we give an exact expression for the non-retarded limit of the Casimir-Polder interaction between an atom and a perfectly-conducting bowl.  
\end{abstract}


\maketitle

\section{Introduction}\label{sec:Section0}
The importance of the Casimir-Polder interaction in cold-atom physics is by now a well established fact and the theory of atom-surface interactions is well developed. There are theoretical tools available, both numerical \cite{Johnson} and analytical \cite{Scheel,Zietal}, for calculating dispersion forces between atoms and polarizable objects for arbitrary geometries and a variety of optical properties of the surfaces. However, practical calculations of Casimir-Polder forces tend to be complicated because, in one way or another, they require solutions to complex electromagnetic scattering problems. In order to make calculations manageable and render final results to be of practical use, simplifying assumptions of some sorts need to be introduced. One possibility is to focus attention on the so-called non-retarded regime of the Casimir-Polder interaction, also called the van der Waals regime. In this regime one assumes the speed of light to be infinite and therefore any scattering problems whose solutions are required for the computation of the Casimir-Polder force become essentially electrostatic problems. This regime is applicable to atoms or molecules whose distance to the surface is much smaller than the wavelength of their dominant dipole transition. Although going to the non-retarded regime somewhat restricts the applicability of the final results, the benefits are enormous; electrostatics is considerably simpler than electrodynamics plus there are numerous scenarios where the non-retarded calculations are adequate and provide accurate estimates of Casimir-Polder forces. For example, the spacings between molecular energy levels are relatively close, so that the wavelengths of dominant dipole transitions in molecules tend to be relatively large. Thus, the interaction between cold molecules and surfaces often falls into the non-retarded regime. Since the mathematical framework of electrostatics is very well investigated, one is able to obtain exact analytic solutions for some non-trivial geometries that bear more resemblance to realistic experimental set-ups than the oversimplified geometries often studied in the literature.

In this paper we respond to the growing need of experimental physicists for simple and easy-to-use expressions enabling efficient calculations of the Casimir-Polder forces in realistic geometries that tend to arise in modern experiments. The question that we think requires urgent attention is this: what is the interaction of an atom with a dielectric surface onto which some conducting structures have been deposited? This is a very common experimental scenario e.g.~with atom chips. Inspired by recent progress in graphene technologies \cite{HexFlakes} one might even imagine an atom interacting with graphene flake or ribbon supported by a dielectric substrate. Such structures are now routinely made and it is just a matter of time until they find their way into cold-atom physics. 

With this motivation in mind, we model a conductor-patched dielectric surface interacting with a neutral atom or molecule by a perfectly reflecting disc supported by a non-dispersive dielectric half-space interacting with a neutral localized quantum system, represented by a fluctuating electric dipole moment. We assume the interaction to be non-retarded and at zero temperature.

In the following we shall find an exact solution to the problem of a perfectly reflecting disc atop a non-dispersive dielectric, which we think is a quite remarkable result and to the best of our knowledge the first of its kind in the literature. In addition we generate a number of other useful results. Most notably we obtain an exact formula for the van der Waals interaction energy of an atom near a perfectly conducting disc. We anticipate that this result will prove useful for current matter-wave experiments where cold molecules are scattered on radially symmetric objects in order to study the diffraction patterns and in particular the so-called Poisson spot, a phenomenon well-known from standard optics. We also derive an expression for the Casimir-Polder force acting on an atom close to the edge of a conducting half-plane deposited on a dielectric substrate. This formula can be applied to the case of an atom near the edge of a conducting ribbon. Furthermore, in the appendix we give an exact expression for the Casimir-Polder interaction between an atom and a perfectly-conducting bowl.

\section{Non-retarded Casimir-Polder potential}\label{sec:Section1}
At close range the Casimir-Polder interaction between a neutral atom and a perfectly reflecting material structure is non-retarded and can be worked out by methods of electrostatics. Taking the atom to be a point-like electric dipole located at $\br_0$, one can show that the interaction energy with a nearby surface can in rectangular coordinates be written as \cite{nonret}
\begin{equation}
\Delta E = \frac{1}{2\epsilon_0}\lim_{\br,\br' \rightarrow\br_0} \sum_{i=1}^3\langle \mu_i^2 \rangle\nabla_i\nabla'_i G_H(\br,\br')\label{eqn:EnergyShift}
\end{equation}
where $G_H(\br,\br')$ is the homogeneous part of the Green's function, to be explained further below. The sum runs over the three components of the electric dipole moment operator $\mu_i$. The complete Green's function $G(\br,\br')$ is the electrostatic potential at $\br$ due to a unit point charge at $\br'$ and satisfies
\begin{equation}
-\nabla^2G(\br,\br')=\delta^{(3)}(\br-\br')\label{eqn:Poisson}
\end{equation}
with appropriate boundary conditions. For a perfect reflector, the boundary conditions on $G(\br,\br')$ are for it to vanish for any $\br$ on the surface of the reflector. For dielectric surfaces Maxwell's equations imply continuity conditions on the gradient of $G(\br,\br')$. The component of $\boldsymbol{\nabla}G(\br,\br')$ that is parallel to the surface of the dielectric is required to be continuous across the interface, whereas the component normal to the surface of the dielectric is continuous when multiplied by the respective dielectric constant $\epsilon(\br)$ on either side of the interface.  The homogeneous part of the potential is the difference between the Green's function $G(\br,\br')$ and its free-space equivalent, i.e.~the potential of the same point charge in free space, 
\begin{equation}
G_H(\br,\br')=G(\br,\br')-\frac{1}{4\pi}\frac{1}{|\br-\br'|}.
\end{equation}
So far the problem is entirely classical and does not involve quantum electrodynamics. The quantum properties of the atom are accounted for in the expectation values of the electric dipole moment operator $\langle\mu^2_i\rangle\equiv\langle j |\mu_i^2|j\rangle$, where $|j\rangle$ denotes the state of the atom, not necessarily its ground state. The difficulty of determining the energy shift (\ref{eqn:EnergyShift}) lies in calculating the Green's function $G(\br,\br')$  for the geometry of interest.

\section{Conducting half-plane on a dielectric substrate \label{sec:Section2}}
\subsection{Green's function}\label{sec:Section2A}
\begin{figure}[b]
\includegraphics[width=7.0 cm, height=7 cm]{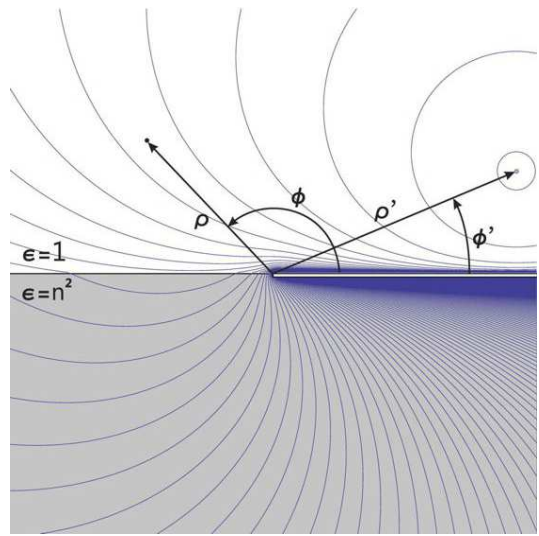}
\caption{\label{fig:geometry} (Color online) A point charge at $\br'=(\rho',\phi',z')$ near a perfectly reflecting half-plane lying on top of a dielectric half-space with constant and frequency independent dielectric function $\epsilon=n^2$ generates the potential given in Eq.~(\ref{eqn:HPGreensFun}). Also shown are lines of constant potential.}
\end{figure}
In this section we derive the Green's function of the Poisson equation for a semi-infinite conducting half-plane lying on top of a dielectric substrate, as depicted in Fig. \ref{fig:geometry}. We work in cylindrical coordinates $\br=(\rho,\phi,z)$ and place the origin of the coordinate system on the edge of the conducting half-plane which is described by  $\phi=0$. The plane $\phi=\pi$ describes the surface of the polarizable body that is not coated by the conductor. We require the electrostatic potential $G(\br,\br')$ to vanish on the surface of the conducting half-plane, that is
\begin{equation}
G(\br,\br')=0\;\; {\rm for}\;\phi=\{0,2\pi\}.
\end{equation}
Across the half-plane described by $\phi=\pi$ we impose standard dielectric continuity conditions,
\begin{equation}
\frac{\partial}{\partial \rho} G(\br,\br'),\;\; \frac{\partial}{\partial z} G(\br,\br'),\;\;\epsilon(\phi)\frac{\partial}{\partial \phi} G(\br,\br')\;\;\;{\rm continuous}.
\label{contcond}
\end{equation}
To find the solution of Eq.~(\ref{eqn:Poisson}) that satisfies the boundary and continuity conditions described above, we work with the eigenfunctions $\Psi_n(\rho,\phi,z)$ of the Laplace operator in cylindrical coordinates and construct the Green's function in terms of an eigenfunction expansion (cf.~e.g. Sect.~III of Ref.~\cite{nonret}).
In the region $0\leq\phi\leq\pi$ eigenfunctions that are regular for small $\rho$ and that vanish on the half-plane $\phi=0$ are
\begin{equation}
\Psi_n(\rho,\phi,z)=\frac{1}{\sqrt{2\pi}}e^{i\kappa z}J_{\frac{m}{2}}(k\rho)\frac{1}{\sqrt{\pi}}\sin\left(\frac{m\phi}{2}\right).
\label{eigen1}
\end{equation}
In the region $\pi\leq\phi\leq2\pi$ those that are regular for small $\rho$ and vanish on the half-plane $\phi=2\pi$ are
\begin{equation}
\Psi_n(\rho,\phi,z)=-\frac{1}{\sqrt{2\pi}}e^{i\kappa z}J_{\frac{m}{2}}(k\rho)\frac{1}{\sqrt{\pi}}\sin\left(2\pi-\frac{m\phi}{2}\right).
\label{eigen2}
\end{equation}
The parameters $\kappa$, $k$, and $m$ in Eqs.~(\ref{eigen1}) and (\ref{eigen2}) would not {\em a priori} need to be the same, but the requirement of continuity of the $\rho$ and $z$ derivatives, according to Eq.~(\ref{contcond}), forces them to be the same. In order to satisfy the simultaneous continuity condition on $\epsilon(\phi){\partial}\Psi_n(\rho,\phi,z)/\partial \phi$ one needs to have either
$$
\left.\sin\left(\frac{m\phi}{2}\right)\right|_{\phi=\pi}=0 
$$
or
$$
\left.\frac{\partial}{\partial \phi} \sin\left(\frac{m\phi}{2}\right)\right|_{\phi=\pi}
=\left.\frac{m}{2} \cos\left(\frac{m\phi}{2}\right)\right|_{\phi=\pi} =0\, .
$$
In the first case, $m$ must therefore be an even integer, and in the second an odd integer. Putting the parts below and above $\phi=\pi$ together with the appropriate relative factors so as to satisfy the continuity conditions at $\phi=\pi$ and then normalizing each eigenfunction, one can proceed to assembling the Green's function from these eigenfunctions and obtains in the region $0\leq\phi\leq\pi$ and $0\leq\phi'\leq\pi$ 
\begin{eqnarray}
G(\br,\br')=\frac{1}{2\pi^2}\sum_{m=1}^{\infty} \int_{-\infty}^{\infty}d\kappa \int_{0}^{\infty}dk \frac{k}{\kappa^2+k^2} e^{i\kappa(z-z')}\nonumber\\
\times J_{m}(k\rho)J_{m}(k\rho') \frac{2n^2}{n^2+1}\sin m\phi\sin m\phi'\nonumber\\
+\frac{1}{2\pi^2}\sum_{m=0}^{\infty} \int_{-\infty}^{\infty}d\kappa \int_{0}^{\infty}dk \frac{k}{\kappa^2+k^2} e^{i\kappa(z-z')}\nonumber\\
\times J_{m+1/2}(k\rho)J_{m+1/2}(k\rho') \frac{2}{n^2+1}\hspace*{10mm}\nonumber\\
\times\sin\left[ \left(m+\frac12\right)\phi\right]\sin \left[\left(m+\frac12\right)\phi'\right],\nonumber\\
\end{eqnarray}
and in the region $\pi\leq\phi\leq2\pi$ and $0\leq\phi'\leq\pi$ 
\begin{eqnarray}
G(\br,\br')=\frac{1}{2\pi^2}\sum_{m=1}^{\infty} \int_{-\infty}^{\infty}d\kappa \int_{0}^{\infty}dk \frac{k}{\kappa^2+k^2} e^{i\kappa(z-z')}\nonumber\\
\times J_{m/2}(k\rho)J_{m/2}(k\rho') \frac{2}{n^2+1}\sin \frac m 2\phi\sin \frac m 2\phi'.\nonumber\\
\end{eqnarray}
Carrying out the integrations, as explained in detail in Sect.~IV of Ref.~\cite{nonret}, one then finds for the Green's function in the case of the source being above the material, that is for $\phi'\in(0,\pi)$, 
\begin{eqnarray}
G(\br, \br')&=&\frac{1}{4\pi(n^2+1)}\left\{\frac{1}{D_-}\left[\frac{n^2}{\epsilon(\phi)}+\frac{2\eta_-}{\pi}\arctan\left(\frac{F_-}{D_-}\right)\right]\right.\nonumber\\
&-&\frac{1}{D_+}\left[\frac{n^2}{\epsilon(\phi)}+\frac{2\eta_+}{\pi}\arctan\left(\frac{F_+}{D_+}\right)\right]\label{eqn:HPGreensFun}
\end{eqnarray}
with
\begin{eqnarray}
\epsilon(\phi)=\left\{
\begin{array}{ll}
1 & {\rm for} \phi \in (0,\pi)\\
n^2 & {\rm for} \phi \in (\pi, 2\pi)
\end{array}
\right.
\end{eqnarray}
and 
\begin{eqnarray}
F_\pm&=&\sqrt{2\rho\rho'[1+\cos(\phi\pm\phi')]}\nonumber\\
D_\pm&=&\sqrt{\rho^2+\rho'^2-2\rho\rho'\cos(\phi\pm\phi')+(z-z')^2}\nonumber\\
\eta_\pm&=&{\rm sgn}\left[\cos\left(\frac{\phi\pm\phi'}{2}\right)\right].
\label{eqn:eta}
\end{eqnarray}

\subsection{Energy shift}\label{sec:Section2B}
\begin{figure}[ht]
\includegraphics[trim= 0cm 1cm 0cm 0cm, clip=true, width=8.0 cm, height=5.0 cm]{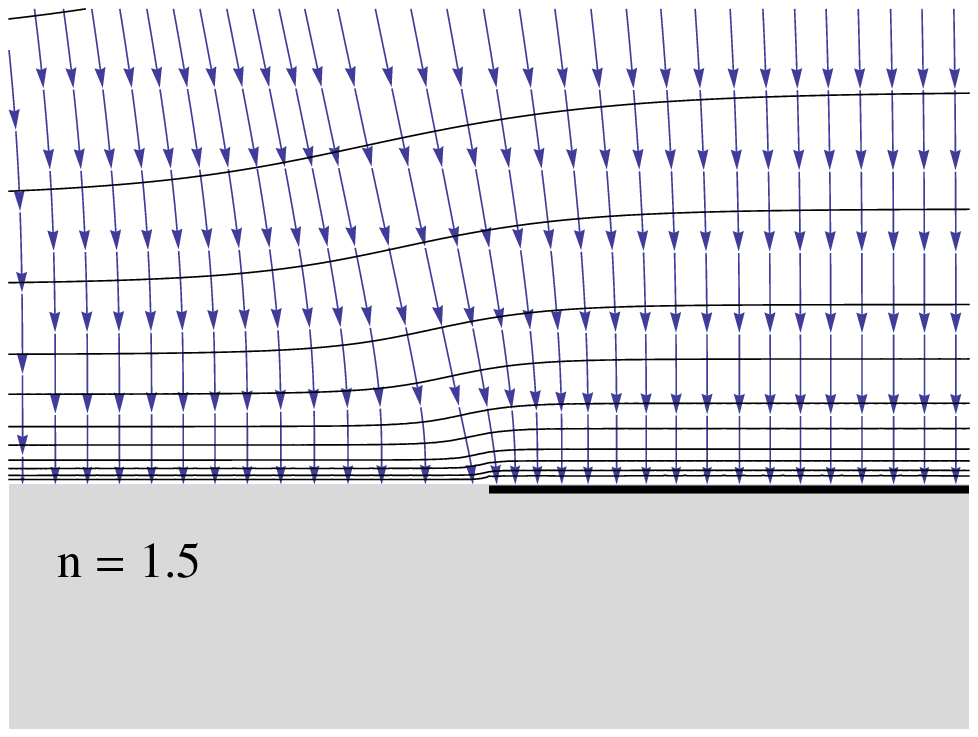}\\
\caption{\label{fig:HSForceIso} (Color online) Direction of the Casimir-Polder force acting on an neutral atom with isotropic polarizability placed near a conducting half-plane lying on a dielectric substrate with refractive index $n=1.5$ (blue arrows). The approximately horizontal lines are contours of constant van der Waals energy.}
\end{figure}
With the Green's function (\ref{eqn:HPGreensFun}) determined, the calculation of the energy shift is now straightforward. We subtract the free-space Green's function from Eq.~(\ref{eqn:HPGreensFun}) and plug the result into formula (\ref{eqn:EnergyShift}). For an atom located at $(\rho,\phi,z)$, cf. Fig. \ref{fig:geometry}, we find that the non-retarded energy shift in the atom may be expressed as
\begin{equation}
\Delta E=-\frac{1}{2\pi^2(n^2+1)\epsilon_0}\left[\Xi_\rho \langle\mu_\rho^2\rangle+\Xi_\phi\langle\mu_\phi^2\rangle+\Xi_z\langle\mu_z^2\rangle\right]
\end{equation}
with the abbreviations
\begin{eqnarray}
\Xi_\rho &=& \frac{5}{48\rho^3}+\frac{\cos\phi}{16\rho^3\sin^2\phi}+\frac{[(n^2+1)\pi-2\phi][1+\sin^2\phi]}{32\rho^3\sin^3\phi},\nonumber\\
\Xi_\phi &=& -\frac{1}{48\rho^3}+\frac{\cos\phi}{8\rho^3\sin^2\phi}+\frac{[(n^2+1)\pi-2\phi][1+\cos^2\phi]}{32\rho^3\sin^3\phi},\nonumber\\
\Xi_z &=& \frac{1}{24\rho^3}+\frac{\cos\phi}{16\rho^3\sin^2\phi}+\frac{[(n^2+1)\pi-2\phi]}{32\rho^3\sin^3\phi}.\nonumber
\end{eqnarray}
In this form the expression for the energy shift facilitates a clear-cut comparison with the result obtained in Ref.~\cite{nonret} for the case of the atom interacting with a half-plane alone i.e. without any substrate present. However, for practical purposes it is much more convenient to express the shift in terms Cartesian coordinates where the directions of the unit vectors, in terms of which the dipole matrix elements $\langle\mu_i^2\rangle$ are expressed, are position-independent. We set up the Cartesian coordinate system in such a way that the $z$ direction is perpendicular to the surface and the $x$ axis runs along the edge of the conducting sheet. Mathematically we set $\rho\cos\phi=y,\;\rho\sin\phi=z,\;z=x$ in Eq.~(\ref{eqn:HPGreensFun}) and calculate the energy shift again by using (\ref{eqn:EnergyShift}). We obtain
\begin{equation}
\Delta E=-\frac{1}{2\pi^2(n^2+1)\epsilon_0}\left[\Xi_x \langle\mu_x^2\rangle+\Xi_y\langle\mu_y^2\rangle+\Xi_z\langle\mu_z^2\rangle\right]
\end{equation}
with the abbreviations
\begin{eqnarray}
\Xi_x&=&\frac{1}{24\left(y^2+z^2\right)^{3/2}}+\frac{y}{16 z^2\left(y^2+z^2\right)}\nonumber\\
&+&\frac{n^2\pi-2\arctan\left(y/z\right)}{32 z^3},\\
\Xi_y&=&\frac{5y^2-z^2}{48\left(y^2+z^2\right)^{5/2}}+\frac{y^3}{16 z^2\left(y^2+z^2\right)^2}\nonumber\\
&+&\frac{n^2\pi-2\arctan\left(y/z\right)}{32 z^3},\\
\Xi_z&=&\frac{5z^2-y^2}{48\left(y^2+z^2\right)^{5/2}}+\frac{2y^3+3yz^2}{16 z^2\left(y^2+z^2\right)^2}\nonumber\\
&+&\frac{n^2\pi-2\arctan\left(y/z\right)}{16 z^3}.
\end{eqnarray}
For an atom with isotropic polarizability, which is the most common case, we get ${\Xi_x\langle\mu_x^2\rangle+\Xi_y\langle\mu_y^2\rangle+\Xi_z\langle\mu_z^2\rangle=\Xi_{\rm iso}\langle\mu^2\rangle}$ with 
\begin{eqnarray}
\Xi_{\rm iso}&=&\frac{1}{8\left(y^2+z^2\right)^{3/2}}+\frac{y}{4 z^2\left(y^2+z^2\right)}\nonumber\\
&+&\frac{n^2\pi-2\arctan\left(y/z\right)}{8 z^3}\label{eqn:HSShiftIsotropic},
\end{eqnarray}
which is a remarkably simple end result. It is straightforward visualize the result by plotting the direction of the Casimir-Polder force, which is done in Fig. \ref{fig:HSForceIso}. As one would expect from physical intuition, there is a lateral component of the Casimir-Polder force due to the presence of the conducting coating. One can easily convince oneself, analytically or numerically, that the lateral component of the Casimir-Polder force pulling the atom towards the edge of the half-plane is dominated by the normal component of the force even for relatively low values of the index of refraction $n$.

\section{Kelvin inversion\label{sec:Section3}}
The Kelvin inversion \cite{Kelvin} is a non-linear coordinate transformation, a reflection of space in a sphere of radius $S$ and centred at $\mathbf{s}$;
it is defined as follows:
\begin{equation}
\boldsymbol{\mathcal{T}}[\br]=
\dfrac{S^2}{|\br-\mathbf{s}|^2}(\br-\mathbf{s})+\mathbf{s}.\label{eqn:Inversion}
\end{equation}
Ref.~\cite{Jeans} gives a very clear overview of the geometrical properties of this transformation. It is of interest here because it preserves solutions of boundary-value problems of the Poisson equation, as we shall explain further below. 
The Green's function of the Poisson equation $G(\br,\br')$ is a function of two variables: the observation point $\br$ and the source point $\br'$. To apply the transformation (\ref{eqn:Inversion}) to the Green's function $G(\br,\br')$ one applies it to both of its arguments. The so transformed Green's function is, when multiplied by an appropriate pre-factor, a solution to the Poisson equation in the transformed geometry, because one has \cite{Footnote,Wermer}
\begin{equation}
-\nabla^2\left[\frac{S^2}{|\br-\mathbf{s}||\br'-\mathbf{s}|}G\left(\boldsymbol{\mathcal{T}}[\br],\boldsymbol{\mathcal{T}}[\br']\right)\right]=\delta^{(3)}(\br-\br').\label{eqn:NewPoisson}
\end{equation}
Therefore the transformation
\begin{equation}
G(\br,\br')\Rightarrow \frac{S^2}{|\br-\mathbf{s}||\br'-\mathbf{s}|} G\left(\boldsymbol{\mathcal{T}}[\br],\boldsymbol{\mathcal{T}}[\br']\right)\equiv \bar{G}(\br,\br')\label{eqn:FullInversion}
\end{equation}
generates a new Green's function of the Poisson equation $\bar{G}(\br,\br')$ in a new geometry from an already known Green's function $G(\br,\br')$ in a geometry that is related to the new one by a Kelvin transformation. It generates a new solution to a boundary-value problem in a new geometry from the known solution in the original geometry, because, if the Green's function $G(\br,\br')$ vanishes on some surface $\sigma $, then the transformed Green's function $\bar{G}(\br,\br')$ vanishes on the transformed surface $\bar{\sigma}$. In this way Green's functions for perfect reflectors of various shapes can be obtained from known Green's functions by various Kelvin transformations, i.e. by adjusting the position and the radius of the inversion sphere \cite{Keijo, Keijo2}. 

For the present purposes it is important to note that, in general, the Kelvin transformation does not preserve dielectric boundary conditions, which is why it is normally thought to be useful only for problems involving just perfect reflectors. 
The most obvious example of a case in which the Kelvin inversion does not preserve the solution of the Poisson equation is that of a dielectric sphere, in contrast to
the potential of a point charge near a perfectly conducting sphere which may be obtained by applying the Kelvin inversion to the potential near the perfectly reflecting flat mirror.
The potential near a flat mirror, perfectly reflecting or dielectric, is a sum of a free-space potential and its image in the mirror. 
The potential near a perfectly reflecting sphere also has the same structure, and the image charge and its location can be obtained from the flat-mirror solution by a Kelvin transformation. However, the potential of a point charge near a dielectric sphere cannot be written down as a sum of a free-space potential plus an image potential from just a single point image charge. The exact solution in this geometry is more complicated and  involves a point image charge and an additional continuous line of image charges inside the sphere \cite{Lindell}. Therefore the solutions in those two geometries, for the dielectric half-space and for the dielectric sphere, cannot be connected by the Kelvin transformation (\ref{eqn:Inversion}). It is easy to understand why this is the case: a dielectric surface enforces different continuity conditions on the gradients of the potential normal and parallel to it, but in general Kelvin transformations distort geometries and hence do not preserve normal and parallel directions. However, there are some special circumstances under which the Kelvin inversion does 
in fact preserve normal and parallel directions at a surface, for example, for a cone whose tip is the centre of the inversion.
This transformation is trivial as it maps a dielectric cone into itself but, as we shall see in the next section, it is nevertheless very useful as it can be applied to cones with conducting coatings. Note that a plane is a cone with $180^\circ$ opening angle, and thus the geometry of Fig. \ref{fig:HSForceIso}, for which the  Green's function is given by Eq.~(\ref{eqn:HPGreensFun}), is a special case of such a cone.

\section{Conducting disc on a dielectric surface}\label{sec:Section4}
\subsection{Green's function}\label{sec:Section3A}
\begin{figure}[ht]
\includegraphics[width=7.0 cm, height=4 cm]{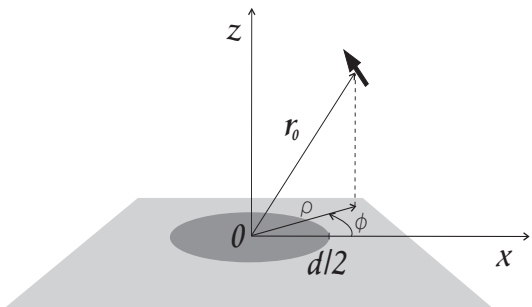}\\
\caption{\label{fig:Disc} (Color online) A conducting disc lying on top of a dielectric substrate, which is a dielectric half-space with dielectric constant $\epsilon=n^2$.}
\end{figure}
In this section we are going to derive the Green's function of the Poisson equation for the geometry of the conducting disc lying on top of a dielectric substrate, cf. Fig \ref{fig:Disc}. We take the perfectly reflecting disc of diameter $d$ to lie in the $z=0$ plane, centred at the origin. The electrostatic potential is required to vanish on the surface of the disc, i.e.
\begin{equation}
G(\br,\br')=0\;\; {\rm for}\;\; z=0\;\;\cap\;\; x^2+y^2 \le \frac{d^2}{4}\; ,
\end{equation}
where $x,y$ are the coordinates in the plane of the disc. In the plane $z=0$ outside the disc we require that $G(\br,\br')$ satisfies dielectric boundary conditions, that is, we require that 
\begin{eqnarray}
\frac{\partial}{\partial x}G(\br,\br'),\;\frac{\partial}{\partial y}G(\br,\br'),\;\epsilon(z)\frac{\partial}{\partial z}G(\br,\br')\nonumber
\end{eqnarray}
are continuous for $z=0\;\cap\; x^2+y^2 \ge d^2/4$.
\begin{figure}[ht]
\includegraphics[width=8.5 cm, height=2.5 cm]{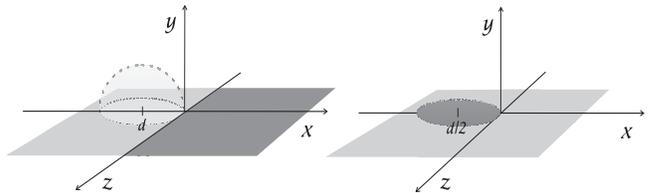}\\
\caption{\label{fig:Inversion} (Color online) Green's function (\ref{eqn:HPGreensFun}) applies to the geometry of a conducting half-plane on a dielectric substrate where the perfect reflector occupies the $y=0\;\cap\;x\ge0$ plane (left). Application of a Kelvin inversion
(\ref{eqn:Inversion}) with $\mathbf{s} = (-d,0,0)$ and $S = d$ yields the Green's function for a conducting disc on a dielectric substrate (right).}
\end{figure}
Obtaining the Green's function with the above boundary and continuity conditions by standard methods, such as an eigenfunction expansion, would be a formidable, if not impossible task. Instead, we are going to derive the required Green's function by means of a Kelvin inversion, as described in the previous section, starting from the Green's function (\ref{eqn:HPGreensFun}) obtained in Sec. \ref{sec:Section2A} for a half-space covered with a conducting half-plane. As explained at the end of Sec. \ref{sec:Section3}, a Kelvin transformation preserves the correct continuity conditions on the surface of the dielectric provided the centre of the inversion sphere is placed somewhere on the surface of the dielectric substrate.
We place the centre of the inversion sphere at $\mathbf{s}=(-d,0,0)$ and set its radius $S=d$. For this particular choice of $\mathbf{s}$ and $S$ the transformation leaves the dielectric substrate unchanged while mapping the conducting half-plane $\sigma=\{\br \in \mathbb{R}^3 : y=0\;\cap\;x\ge 0\}$, cf. Fig. \ref{fig:Inversion} (left), into a disk $\bar{\sigma}=\{\br\in \mathbb{ R}^3:y=0\;\cap\; (x+d/2)^2+z^2 \le (d/2)^2\}$, cf. Fig. \ref{fig:Inversion} (right). This ensures the correct boundary and continuity conditions across the whole surface $y=0$. 

To simplify the notation, we wish to center the coordinate system on the disc and orient the $z$-axis perpendicular to the sheet, as shown in Fig. \ref{fig:Disc}. Thus we shift and rotate the axes according to $x\rightarrow y-d/2,\; y\rightarrow z,\; z\rightarrow x$. We are interested in the disc-geometry Green's function for the case $z,z'>0$, so that we need the Green's function (\ref{eqn:HPGreensFun}) for the case $\phi,\phi'\in (0,\pi)$. In that case Eq.~(\ref{eqn:eta}) reads
\begin{eqnarray}
\eta_-=1,\;\;\;
\eta_+={\rm sgn}\left[\sin\left(\phi+\phi'\right)\right].
\end{eqnarray}
It is advantageous to further rewrite $\eta_+$ in terms of ${\sin (\phi+\phi')=\sin(\phi)\cos(\phi')+\cos(\phi)\sin(\phi')}$ as this facilitates the straightforward application of the Kelvin inversion. The procedure described above then yields the transformed Green's function in the same form as in Eq. (\ref{eqn:HPGreensFun}) but with
\begin{eqnarray}
F_\mp &=&\frac{\sqrt{2}}{d}\left\{\left(\rho^2+z^2-\frac{d^2}{4}\right)\left(\rho'^2+z'^2-\frac{d^2}{4}\right)\right.\nonumber\\
& \pm &
d^2zz'+
\sqrt{
\bigg[z^2+\left(\rho-\frac{d}{2}\right)^2\bigg]
\bigg[z^2+\left(\rho+\frac{d}{2}\right)^2\bigg]
}
\nonumber\\
&\times &\left.
\sqrt{
\bigg[z'^2+\left(\rho'-\frac{d}{2}\right)^2\bigg]
\bigg[z'^2+\left(\rho'+\frac{d}{2}\right)^2\bigg]
}
\right\}^{1/2},\label{eqn:NewF}\\
D_\mp &=& \sqrt{\rho^2+\rho'^2-2\rho\rho'\cos(\phi-\phi')+(z\mp z')^2},\label{eqn:NewD}\\
\eta_+ &=& {\rm sgn}[z(d^2/4-\rho'^2-z'^2)+z'(d^2/4-\rho^2-z^2)],\;\;\;\;\;\;\label{eqn:NewEta}
\end{eqnarray}
and, of course, $\eta_-=1$. We emphasize that this result, i.e. Eq. (\ref{eqn:HPGreensFun}) with the insertions (\ref{eqn:NewF})--(\ref{eqn:NewEta}), is valid for $z,z'>0$, when the source and observation points are both in the empty space above the substrate, cf. Fig \ref{fig:Disc}. 
The Green's function for the case of the source in vacuum but the observation point inside the material can also be obtained quite easily by considering $\eta_{\pm}$ in Eq.~(\ref{eqn:eta}) and making appropriate amendments.

\subsection{Energy shift}\label{sec:Section3B}
\begin{figure}[ht]
\includegraphics[trim= 3cm .2cm 3cm 3cm, clip=true, width=8.0 cm, height=4 cm]{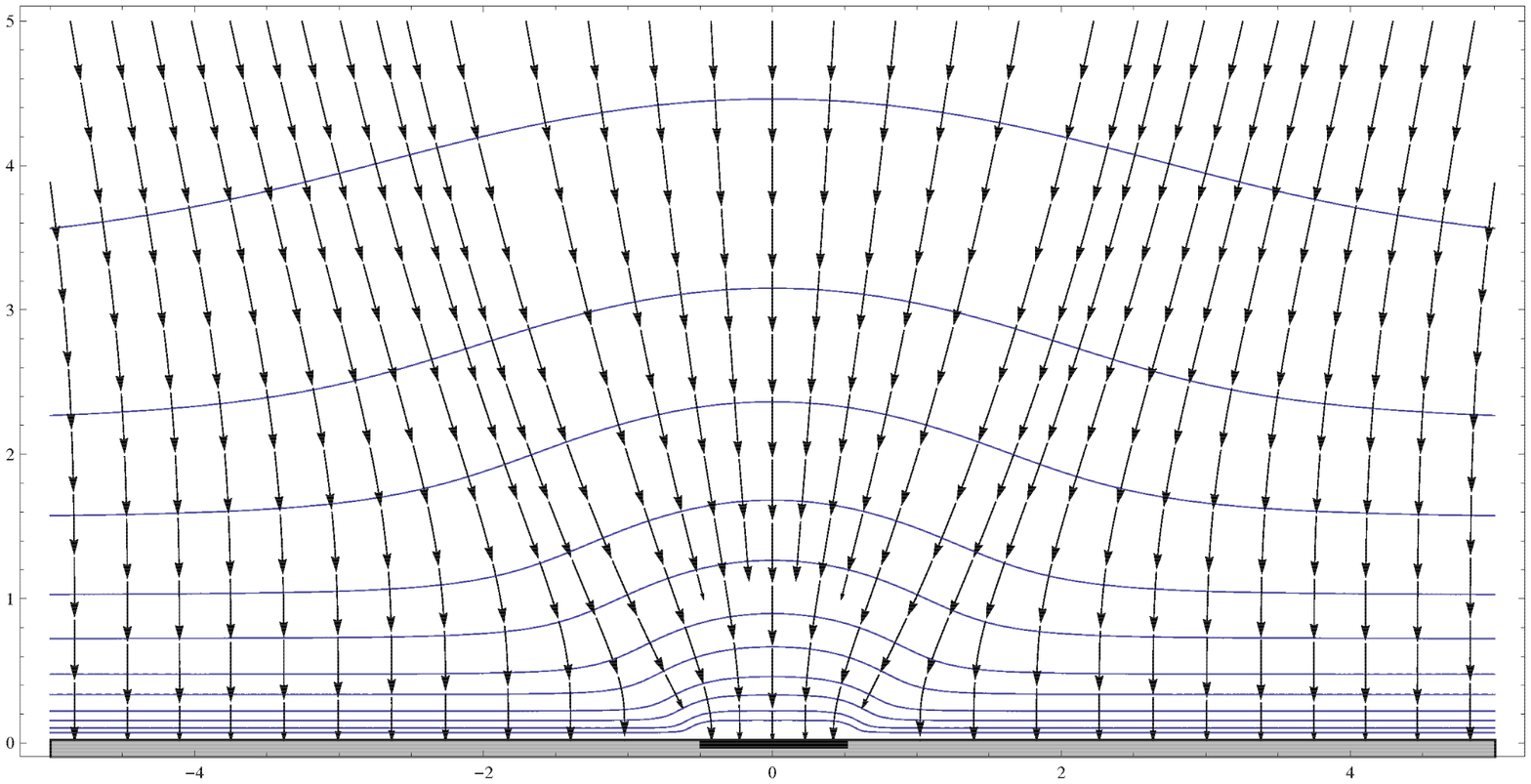}\\
\caption{\label{fig:Disc_On_HS_Force} (Color online) Direction of the Casimir-Polder force acting on a neutral atom with isotropic polarizability placed near a conducting disc lying atop a dielectric substrate of refractive index $n=1.5$. The approximately horizontal lines are contours of constant van der Waals energy.}
\end{figure}
Knowing the Green's function we can now calculate the energy shift. We subtract the free-space potential from the result for the Green's function that we obtained in the Sec. \ref{sec:Section3B} and substitute it in into formula (\ref{eqn:EnergyShift}). For an atom located at $(\rho,\phi,z)$, cf. Fig. \ref{fig:Disc}, we find that the non-retarded energy shift may be expressed as
\begin{equation}
\Delta E=-\frac{1}{8\pi^2(n^2+1)\epsilon_0}\left[\Xi_\rho \langle\mu_\rho^2\rangle+\Xi_\phi\langle\mu_\phi^2\rangle+\Xi_z\langle\mu_z^2\rangle\right]\label{eqn:DiscHSShift}
\end{equation}
with the abbreviations
\begin{eqnarray}
\Xi_\rho&=&\frac{d}{R_+^3R_-^3}\left(\frac{d^2}{6}-\rho^2\right)+\frac{2d\rho^2}{R_+^4R_-^4}\left(\rho^2+z^2-\frac{d^2}{4}\right)\nonumber\\
&+&\frac{2d\rho^2}{R_+^5R_-^5}\left(\rho^2+z^2-\frac{d^2}{4}\right)^2\nonumber\\
&+&\frac{1}{4z^3}\left\{\arctan\left(\frac{d^2/4-\rho^2-z^2}{dz}\right)+\frac{\pi}{2}n^2\right.\nonumber\\
&-&\left.\frac{dz}{R_+^4R_-^4}\left(\rho^2+z^2-\frac{d^2}{4}\right)\right. \nonumber\\
&\times & \left.\left[\left(\frac{d^2}{4}+z^2-\rho^2\right)^2+8z^2\rho^2\right]\right\}
\label{eqn:DiscXiRho}\\
\Xi_\phi&=&\frac{d^3/6}{R_+^3R_-^3}+\frac{1}{4z^3}\left[\arctan\left(\frac{d^2/4-\rho^2-z^2}{dz}\right)+\frac{\pi}{2}n^2\right.\nonumber\\
&-&\left.\frac{dz}{R_+^2R_-^2}\left(\rho^2+z^2-\frac{d^2}{4}\right)\right]\label{eqn:DiscXiPhi}\\
\Xi_z&=&-\frac{d}{R_+^3R_-^3}\left(\frac{d^2}{12}+z^2\right)+\frac{2dz^2}{R_+^5R_-^5}\left(\rho^2+z^2+\frac{d^2}{4}\right)^2\nonumber\\
&-&\frac{d\rho^2}{R_+^4R_-^4}\left(\rho^2+z^2-\frac{d^2}{4}\right)\nonumber\\
&+&\frac{1}{2z^3}\left[\arctan\left(\frac{d^2/4-\rho^2-z^2}{dz}\right)+\frac{\pi}{2}n^2\right.\nonumber\\
&+&\left.\frac{dz}{R_+^2R_-^2}\left(\frac{d^2}{4}-\rho^2+z^2 \right)\right],\label{eqn:DiscXiZ}
\end{eqnarray}
where we have defined $R_{\pm}=\sqrt{(\rho\pm d/2)^2+z^2}$. The result (\ref{eqn:DiscHSShift}) together with Eqs. (\ref{eqn:DiscXiRho})--(\ref{eqn:DiscXiZ}) is, as far as electrostatics is concerned, exact. It applies to atoms whose distance from the surface is much smaller than the wavelength of their dominant dipole transition, so that effects of retardation are unimportant. The simplicity of the analytic expressions in Eqs. (\ref{eqn:DiscXiRho})--(\ref{eqn:DiscXiZ}) makes it easy to evaluate and plot the non-retarded Casimir-Polder force felt by an atom due to the presence of a dielectric substrate with a conducting circular patch. In Fig. \ref{fig:Disc_On_HS_Force} we show the direction of the force acting on an atom with isotropic polarizability. As expected there is a lateral component in the Casimir-Polder force that drags the atom towards the conducting patch.
\begin{figure}[ht]
\includegraphics[trim= 3cm .6cm 3cm 3cm, clip=true, width=8.0 cm, height=4 cm]{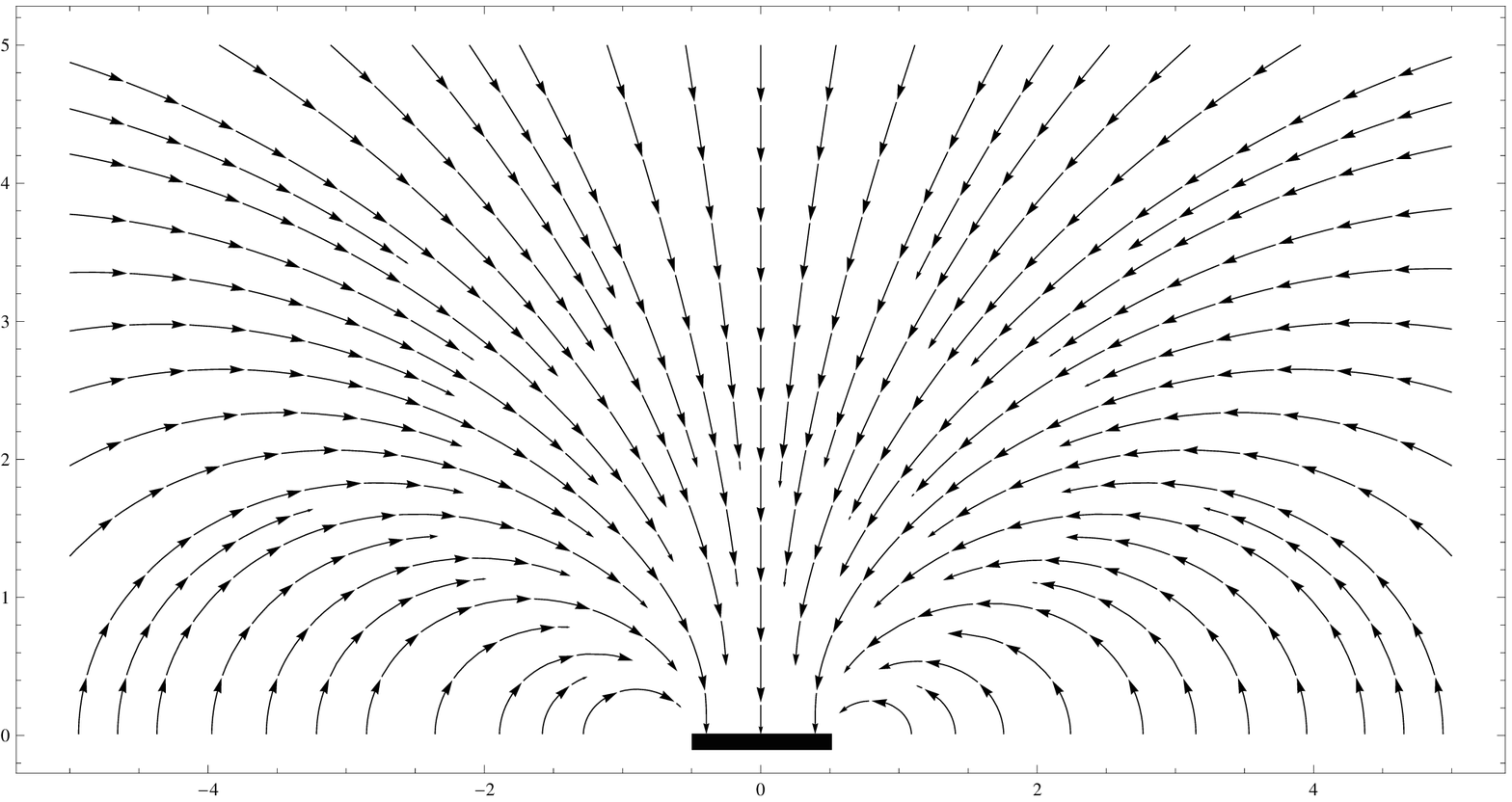}\\
\caption{\label{fig:Disc_Force_Z} (Color online) Direction of the Casimir-Polder force acting on a neutral atom placed near a conducting disc. The atom is polarized in the direction perpendicular to the surface of the disc. }
\end{figure}

Furthermore, it is interesting to consider an atom polarized in the $z$ direction interacting with just a disc, i.e. without any substrate. To this end we take $\langle\mu_\rho^2\rangle=\langle\mu_\phi^2\rangle =0$ in Eq. (\ref{eqn:DiscHSShift}), so that the whole of the energy shift comes from the function $\Xi_z$, and take the limit of no substrate, $n\rightarrow 1$. For such an atom that is polarized in the $z$ direction the energy shift vanishes in the plane of the disc, and similarly to the interaction of an atom with a conducting sheet with a circular hole, as studied in Ref.~\cite{PlateWithHole}, we expect some intricate behaviour of the Casimir-Polder force in this case. 
Since our results are exact we can easily visualize the force in such a situation, which is what we do in Fig.~\ref{fig:Disc_Force_Z}. We note that along the edge of the disc there is a region where the component of the Casimir-Polder force normal to the surface of the disc points away from the disc. This unusual behaviour of the Casimir-Polder force may potentially have an impact on matter-wave experiments which are sensitive to van der Waals interactions. For example, Ref.~\cite{PoissonSpot} studies the interference patterns (Poisson spot) from molecules scattered by a disc. The theoretical calculations used in conjunction with such matter-wave experiments tend to rely on rather oversimplified expressions for the atom-obstacle dispersive interactions \cite{BroglieInterferometry}. As we can see in Fig. \ref{fig:Disc_Force_Z}, the Casimir-Polder interaction for a polarized molecule near a disc exhibits a complicated pattern,  so that theoretical predictions using simplistic models may potentially differ appreciably from the true values that can be calculated on the basis of the exact results derived here, especially for beams of polarized molecules. Although Eqs. (\ref{eqn:DiscXiRho})--(\ref{eqn:DiscXiZ}) may seem complicated at first glance they are expressed in terms of only elementary functions and can thus be evaluated numerically without much effort.

\section{Conclusions \label{sec:Section4}}
\begin{figure}[ht]
\includegraphics[trim= 1cm 1cm 1cm 1cm, clip=true, width=8.0 cm, height=8 cm]{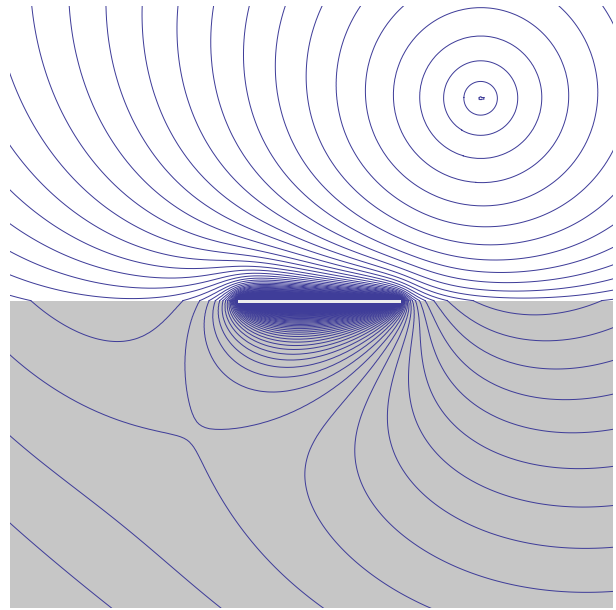}\\
\caption{\label{fig:DiscPotential} (Color online) Contours of constant electrostatic potential $\phi(\br)$ for a point charge above a dielectric half-space with a perfectly conducting disc on top. The index of refraction of the substrate is $n=2$.}
\end{figure}

We have fulfilled our aim of calculating the exact van der Waals potential for a neutral atom interacting with a conducting disc supported by a dielectric half-space. To obtain the exact solution for this complicated but practically very relevant geometry we first solved the Poisson equation for the geometry of a conducting half-plane that is supported by a dielectric half-space, cf. Sec \ref{sec:Section2A}. Then, in Sec. \ref{sec:Section3A} we applied a Kelvin inversion to obtain the Green's function for a conducting circular disc on a dielectric substrate. The result for this Green's function is exact and, to the best of our knowledge, the first of its kind in the literature. It can be used to calculate static electric fields near dielectric surfaces with conducting coatings or patches, cf. Fig \ref{fig:DiscPotential}. The main goal of this paper was to determine the energy-level shift (van der Waals energy) for an atom in close proximity of such a structure. The remarkably simple result for the shift given by Eq.~(\ref{eqn:DiscHSShift}) and Eqs.~(\ref{eqn:DiscXiRho})--(\ref{eqn:DiscXiZ}) allows one to estimate the Casimir-Polder interaction for atoms near dielectric substrates with conducting structures on top. One could imagine an electroplated and partially etched surface or a graphene flake on a ${\rm SiO}_2$ substrate or a graphene quantum dot. In the limit of no substrate our result reduces to that for a flat disc. Detailed analysis of the atom-disc interaction reveals the intricate behaviour of the Casimir-Polder force acting on atoms or molecules polarized in the direction normal to the surface of the disc, cf. Fig. \ref{fig:Disc_Force_Z}. This, as we pointed out at the end of Sec. \ref{sec:Section3B}, may be of significance for  matter-wave interferometry experiments. 

In Appendix \ref{sec:Section5} we obtain the exact expression for the interaction of a neutral atom with a conducting spherical bowl. Our interest in this geometry was ignited by nanocups, i.e. dielectric spheres that are half-coated in gold, which have been recently shown to enable second harmonic generation \cite{nanocup}. However, our result does not include the dielectric filling of the nanocup and is therefore applicable only approximately for weak dielectric substrates.

\appendix
\section{Conducting spherical shell. Green's function and the energy shift \label{sec:Section5}}
\begin{figure}[ht]
\includegraphics[width=8.0 cm, height=3.8 cm]{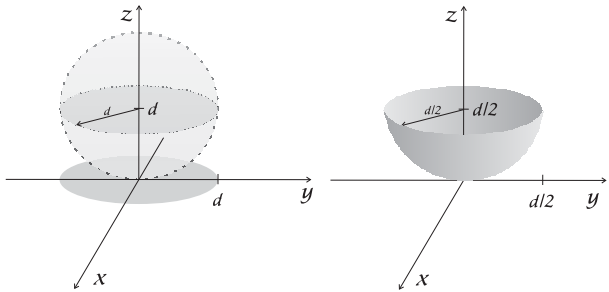}\\
\caption{\label{fig:Disc_To_Shell} (Color online) Applying a suitable Kelvin inversion (\ref{eqn:Inversion}) to the Green's function for a disc generates the Green's function for a conducting spherical bowl. Note that dielectric continuity conditions are not preserved by this transformation, so that the procedure works only in the absence of the substrate, i.e. in the limit $n\rightarrow 1$.}
\end{figure}

\begin{figure}[ht]
\includegraphics[trim= 1cm 1cm 1cm 1cm, clip=true, width=8.0 cm, height=8 cm]{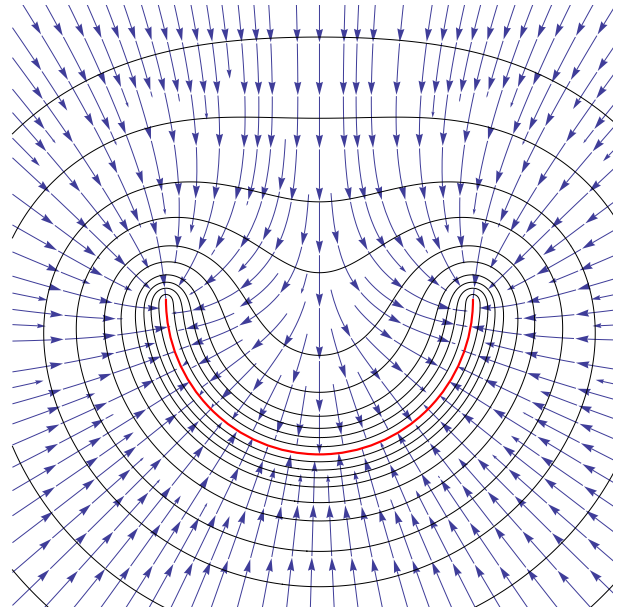}\\
\caption{\label{fig:Shell_Iso} (Color online) Direction of the Casimir-Polder force acting on an atom with isotropic polarizability placed near a conducting spherical bowl (blue arrows). The black circular lines are contours of constant van-der-Waals energy, and the red semicircle is the conducting bowl.}
\end{figure}

The Green's function for a conducting disc without a substrate facilitates the derivation of yet another interesting exact solution in electrostatics, namely the potential due to a point charge in the presence of a conducting spherical bowl. In order to derive the Green's function in this geometry we apply a Kelvin inversion to the Green's function for a conducting disc, which is obtained by taking the limit $n\rightarrow 1$ in the results of Sec. \ref{sec:Section3A}. We place the disc of radius $d$ in the $z=0$ plane and choose $\mathbf{s}=(0,0,d)$ as the centre of the Kelvin inversion and $S=d$ as the radius of the inversion sphere, cf. Eq. (\ref{eqn:Inversion}) and Fig. \ref{fig:Disc_To_Shell}. 
We dispense with the details of the calculation and explicit formulae for the Green's function, which have already been reported in Ref.~\cite{Keijo}. Here we are interested only in giving the energy-level shift for an atom interacting with a conducting spherical bowl. We take the centre of the bowl to be at the centre of the Cartesian coordinate system so that its surface is described by
\begin{equation}
x^2+y^2+z^2=\frac{d^2}{4}\;\;\cap\;\; z\le 0\;.
\end{equation}
Then the energy shift can be expressed in Cartesian coordinates as
\begin{equation}
\Delta E=\frac{1}{16\pi^2\epsilon_0}\left[\Xi_x \langle\mu_x^2\rangle+\Xi_y\langle\mu_y^2\rangle+\Xi_z\langle\mu_z^2\rangle\right]
\end{equation}
with the abbreviations
\begin{eqnarray}
\Xi_x&=&\frac{d^2z}{s^4+d^2z^2}\left( \frac{x^2}{s^4+d^2z^2}+\frac{x^2+d^2/4}{s^4}\right)\nonumber\\
&-&\frac{2dx^2s^4}{\left(s^4+d^2z^2\right)^{5/2}}-\frac{d}{\left( s^4+d^2z^2\right)^{3/2}}\left(\frac{d^2}{6}-x^2\right)\nonumber\\
&+&\frac{d\left(x^2+d^2/4\right)}{s^6}\left[\arctan\left(\frac{dz}{|s^2|}\right)-\frac{\pi}{2}\right]\label{eqn:ShellX}\\
\Xi_z&=&\frac{d^2z}{s^4+d^2z^2}\left[ \frac{z^2-d^2/4}{s^4+d^2z^2}+\frac{d^2/2-x^2-y^2}{s^4}\right.\nonumber\\
&+&\left.\frac{d^2z^2}{s^2\left(s^4+d^2z^2\right)}\right]-\frac{2dz^2\left(s^2+d^2/4\right)^2}{\left(s^4+d^2z^2\right)^{5/2}}\nonumber\\
&+&\frac{d}{\left( s^4+d^2z^2\right)^{3/2}}\left(\frac{d^2}{12}+z^2\right)\nonumber\\
&+&\frac{d\left(z^2+d^2/4\right)}{s^6}\left[\arctan\left(\frac{dz}{|s^2|}\right)-\frac{\pi}{2}\right]\;.\label{eqn:ShellZ}
\end{eqnarray}
Here $s^2=x^2+y^2+z^2-d^2/4$, and the result for $\Xi_y$ is obtained by the replacement $x \leftrightarrow y$, that is, $\Xi_y(x,y,z,d)=\Xi_x(y,x,z,d)$. It may at first sight seem awkward to work in Cartesian coordinates in this geometry, but this in fact avoids mathematical difficulties which arise at the origin of the spherical coordinate system where the azimuthal angle is not well defined. Besides, it is useful to have dipole matrix elements expressed in terms of unit vectors whose orientation does not depend on their position. We plot the direction of the Casimir-Polder force acting on an atom with isotropic polarizability in Fig. \ref{fig:Shell_Iso}.

\begin{acknowledgments} 
We would like to acknowledge financial support from the UK Engineering and
Physical Sciences Research Council.
\end{acknowledgments}


\begin{thebibliography}{99}
\bibitem{Footnote}{The RHS of Eq. (\ref{eqn:NewPoisson}) contains an additional term proportional to $\delta^{(3)}(\br-\mathbf{s})\;G\left(\boldsymbol{\mathcal{T}}[\br],\boldsymbol{\mathcal{T}}[\br']\right)$, which vanishes provided $G\left(\boldsymbol{\mathcal{T}}[\br],\boldsymbol{\mathcal{T}}[\br']\right)$ vanishes for $\boldsymbol{\mathcal{T}}[\br]\rightarrow\infty$.}
\end{thebibliography}
\end{document}